\documentclass[a4paper]{article}

\usepackage{amsmath,amsthm,amssymb,mathtools}
\usepackage{comment}
\usepackage[T1]{fontenc} 
\usepackage[mathcal]{euscript}
\usepackage{graphicx}
\usepackage[hidelinks]{hyperref}
\usepackage[top=3.2cm,bottom=3.2cm,left=2.98cm,right=2.98cm]{geometry}
\usepackage{tikz}
\usepackage[export]{adjustbox}
\usepackage{appendix}
\usepackage{setspace}
\usepackage{tcolorbox}
\numberwithin{equation}{section}
\theoremstyle{plain}

\theoremstyle{definition}

\makeatletter
\g@addto@macro\bfseries{\boldmath}
\makeatother

\usetikzlibrary{shapes.geometric,decorations.markings,intersections,calc,positioning}

\tikzset{->-/.style={decoration={
  markings,
  mark=at position .6 with {\arrow{>}}},postaction={decorate}}}

\tikzset{-<-/.style={decoration={
  markings,
  mark=at position .6 with {\arrow{<}}},postaction={decorate}}}

\tikzset{%
    add/.style args={#1 and #2}{
        to path={%
 ($(\tikztostart)!-#1!(\tikztotarget)$)--($(\tikztotarget)!-#2!(\tikztostart)$)%
  \tikztonodes},add/.default={.2 and .2}}
}  

\tikzset{
    extended line/.style={shorten >=-#1,shorten <=-#1},
    extended line/.default=1cm]
}

\tikzset{line through/.style args={#1 parallel to line through #2 and #3 and
length #4}{insert path={%
let \p1=($(#3)-(#2)$),\n1={atan2(\y1,\x1)} in (#1) -- ++ (\n1:#4)}}}

\definecolor{col1}{rgb}{0.4, 0.69, 0.2}
\definecolor{col2}{rgb}{0.96, 0.29, 0.54}

\definecolor{green(ryb)}{rgb}{0.4, 0.69, 0.2}
\definecolor{frenchrose}{rgb}{0.96, 0.29, 0.54}
\definecolor{persianblue}{rgb}{0.11, 0.22, 0.73}
\definecolor{jade}{rgb}{0.0, 0.66, 0.42}
\definecolor{limegreen}{rgb}{0.2, 0.8, 0.2}

\usepackage{graphicx}

\title{Iterative polynomial regularisation  of  $\text{P}_{\text{IV}}$-type: \\ Hamiltonian systems and 
Newton polygons}
\author{Marta Dell'Atti{\color{jade}$\,^1$} \\[-.5ex] \small{\url{m.dell-atti@uw.edu.pl}} \\[1ex] Galina Filipuk{\color{jade}$\,^1$} \\[-.5ex]  \small{\url{g.filipuk@uw.edu.pl}} \\[.5ex]
{\color{jade}$\,^1$}{ \small  University of Warsaw, Institute of Mathematics, Banacha 2, 
Warsaw, 02-097, Poland}
}
\date{}

\begin{document}

\maketitle

\begin{abstract}
We study several polynomial Hamiltonian systems of $\text{P}_{\text{IV}}$-type (including the mixed case quasi-$\text{P}_{\text{IV}}$), and show that via the iterative process of polynomial regularisation, it is possible to identify the ``minimal'' Hamiltonian system. The selected Hamiltonian function is associated with the Newton polygon with minimal area and smallest highest total degree. 
\end{abstract}

\section{Introduction}
\setlength{\belowdisplayskip}{6.2pt} \setlength{\belowdisplayshortskip}{6.2pt}
\setlength{\abovedisplayskip}{6.2pt} \setlength{\abovedisplayshortskip}{6.2pt}

\pgfmathsetmacro\MathAxis{height("$\vcenter{}$")}

The famous six Painlev\'e equations are nonlinear second order differential equations of type 
\begin{equation}\label{eq:sec_order}
    x''=R(x', x;z) \,, \qquad z \in \mathbb{C}\,, 
\end{equation}
with $R$ rational in the first two arguments, whose solutions $x(z)$ are not expressible in terms of known functions~\cite{Painleve1900,Fuchs1907,Gam}. These equations appear in many areas of mathematics and mathematical physics and have numerous fascinating properties, among which the characterisation of their singularities. 
In particular, the nonlinearity of~\eqref{eq:sec_order} induces the occurrence of (infinitely many) movable singularities $z_*$, whose location depends on the initial conditions. The Painlev\'e equations satisfy the Painlev\'e property, for which the only admissible movable singularities are poles~\cite{JosKru,Shimomura2003}. 
Their solution $x(z)$ can thus be locally expanded into the Laurent series in $z-z_*$,~i.e. 
\begin{equation}\label{eq:laurent}
    x(z)=\sum_{j=\ell}^{\infty} a_{j}\,(z-z_*)^j\,, \qquad \ell \in \mathbb{Z}\,,
\end{equation}
with the finite principal part in the neighbourhood of the movable singularity $z_*$, and $|\ell|$ the ramification number~\cite{IwasakiOkada}.   

An interesting extension of the Painlev\'e property is the so-called quasi-Painlev\'e property~\cite{Shimomura2006,Shimomura2008}, which means that solutions admit movable singularities as algebraic branch points, after a finite-length path continuation in the complex plane. In this case, solutions can be expanded in terms of Puiseux series
\begin{equation}\label{eq:puiseux}
    x(z)=\sum_{j=\ell}^{\infty} b_{j}\,(z-z_*)^{j/k}\,, \qquad \ell \in \mathbb{Z},~~ k \in \mathbb{N}\,, 
\end{equation} 
i.e.\ expansions in fractional powers of $z-z_*$ with a finite principal part (see~\cite{TKGF,GF_AS1,GF_AS2,MDTK1,MDTK2} and the numerous references therein). We call $1/k$ in~\eqref{eq:puiseux} the branching order of the expansion, so that for~\eqref{eq:laurent} we have branching order equal to $1$. In this sense, the Laurent series can be seen as a particular case of the Puiseux series. 


This paper focuses on systems associated with the fourth Painlev\'e equation, motivated by observations related for the second Painlev\'e equation in~\cite{GF}. 
The second and fourth Painlev\'e equations are 
\vspace*{-2ex}
\begin{align}
\label{P2}
\text{P}_{\text{II}}\colon& \quad  x''=2x^3 + z\,x+\alpha,\\[1ex]
\label{P4}
\text{P}_{\text{IV}}\colon& \quad  x''=\frac{(x')^2}{2x}+\frac{3}{2}x^3 +4zx^2+2(z^2 -\alpha)x+\frac{\beta}{x},
\end{align}
where $\alpha, \beta \in \mathbb{C}$ are arbitrary parameters, and for some specific values the solutions can be expressed in terms of known classical functions. The Painlev\'e equations admit a Hamiltonian formulation~(see the classical papers by Okamoto~\cite{okamoto1,okamoto2,okamoto3,okamoto4}), i.e.\ there exists a (non-autonomous) function $H(x,y;z)$ in two variables such that Painlev\'e equations are equivalent to the Hamiltonian system 
\begin{equation}\label{Hamsyst}
x'=\frac{\partial H}{\partial y}, \qquad y'=-\frac{\partial H}{\partial x}\,, \qquad \big(x(z), y(z)\big) \in \mathbb{C}^2\,,
\end{equation}
where $'\equiv d/dz$. For $\text{P}_{\text{II}}$ and $\text{P}_{\text{IV}}$ the so-called Okamoto polynomial Hamiltonians are
\begin{align}
\label{OkHamP2}
H_{\text{II}}^{\text{Ok}}\big(x(z),y(z);z\big)&=\frac{y^2}{2}-\left(x^2+\frac{z}{2}\right)y-\left(\alpha+\frac{1}{2}\right)x\,,  \\[1ex]
\label{OkHamP4}
H_{\text{IV}}^{\text{Ok}}\big(x(z),y(z);z\big)&=2x y^2-(x^2 +2zx+2\kappa_0)y+\vartheta_{\infty}x\,,
\end{align}
where $\kappa_0\,, \vartheta_\infty$ are constants related to $\alpha, \beta$ in $\text{P}_{\text{IV}}$ as equation~\eqref{P4} via $\alpha=2\vartheta_{\infty}-\kappa_0+1,\;\beta=-2\kappa_0^2$. 

In this paper, we will associate with the Hamiltonian function its coefficient matrix and the corresponding Newton polygon, so with a slight abuse of notation we will write e.g.\ for~\eqref{OkHamP4}
\vspace*{-2ex}
\begin{equation}\label{eq:HamOkamoto1}
 H_{\text{IV}}^{\text{Ok}}=\left(
\begin{array}{ccc}
 0 & -2 \kappa_0 & 0 \\
  \vartheta_{\infty} & -2 z & 2 \\
 0 & -1 & 0 \\
\end{array}
\right) \hspace{10ex} \includegraphics[width=.16\textwidth,valign=c]{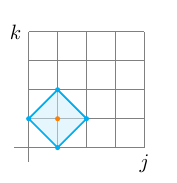} 
\end{equation}
\vspace*{-2ex}

\noindent 
which means that the Hamiltonian function \eqref{OkHamP4} is given by multiplying the vector $(1,x,x^2)$ to the left of the matrix \eqref{eq:HamOkamoto1}, and the transposed vector $(1,y,y^2)^{\top}$ to the right. We follow the construction in~\cite{MDTK2} to depict the Newton polygon: we take the convex hull of all points with non-negative  integer coordinates~$(i,j)$ such that the term $x^i y^j$ with a non-zero coefficient appears in the Hamiltonian\footnote{In general the Hamiltonian function is defined up to a constant term, that we omit in the majority of cases, because we are looking for minimising the areas of the polygons.}. If there are $n$ integer lattice points inside the convex hull, then we say that the Newton polygon has genus $n$, so for instance in the case~\eqref{eq:HamOkamoto1} we have genus $1$. We can also calculate the area of the convex hull, and for the Hamiltonian  
 \eqref{eq:HamOkamoto1} we have the diamond-shaped convex hull and area equal to 2 in lattice units. 
 The bold points in the lattice are the points~$(i,j)$ such that the term $x^i y^j$ with a non-zero coefficient appears in the corresponding Hamiltonian, and the points in orange are other integer lattice points for reader's convenience so that one can easily count them. Additionally, we consider the highest total degree in the Hamiltonian, which is identified as the largest diagonal in the lattice on which (at least one of) the boundary points of the polygon lay, given by $\text{max}(i+j)$ for $(i,j)$ in the polygon. For instance, in~\eqref{eq:HamOkamoto1} this value is $3$.


In this paper, we shall focus on different Hamiltonian systems related to equations of the Painlev\'e~IV-type to find Hamiltonians whose Newton polygons have minimal area and smallest 
total degree. We will consider a generalisation of the Okamoto Hamiltonian~\eqref{OkHamP4}, where we include arbitrary coefficient functions (see equation~\eqref{eq:Ok1}), and a modified version of it with an additional term (see equation~\eqref{eq:Ok2}). Moreover, we will look at the so-called mixed case of quasi-$\text{P}_{\text{IV}}$ by studying two different Hamiltonians (see equations~\eqref{H1 mixed} and~\eqref{H2 mixed}). The term ``mixed case'' refers to the diversified nature of the behaviour of the solution around the movable singularities, e.g.\ for quasi-$\text{P}_{\text{IV}}$ one simple pole, and two algebraic poles of square root type. 

The technique we use throughout the paper is the (iterative) polynomial regularisation of Hamiltonian systems, whose main ideas are summarised in appendix~\ref{app:regularisation}. In the regularisation procedure, the input is a Hamiltonian system in the projective complex space with points of indeterminacy, and the output is given by several regularised Hamiltonian systems, each of which encodes the behaviour of the solution around a specific movable singularity, of the form either~\eqref{eq:laurent} or~\eqref{eq:puiseux}. The final systems are all birationally equivalent to the original one via chains of transformations of coordinates (we shall omit some steps in these computations). We are going to analyse the final systems in terms of the coefficient matrix of the associated Hamiltonian and the related Newton polygon, with particular attention to its area. 
If needed, this procedure will be iterated, promoting each of the above mentioned final systems to be the input of the next regularisation. We shall use the notation $r_k$ for $k=0\,, 1, 2,\ldots$ to label each regularisation, e.g.\ the initial regularisation will be identified by $r_0$ and then subsequent (polynomial) regularisations by $r_k$ with $k>0$. 

Our primary motivation builds upon the observation in~\cite{GF}, where starting from the non-Okamoto polynomial Hamiltonian related to $\text{P}_{\text{II}}$
\vspace*{-3ex}
\begin{equation}\label{eq:P2_nonOka}
H_{\text{II}}\big(x(z),y(z);z\big)=\frac{1}{2}(y^2-x^4-z x^2)-\alpha x  \hspace{10ex} \includegraphics[width=.16\textwidth,valign=c]{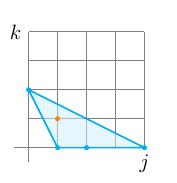} 
\end{equation}
\vspace*{-3ex}

\noindent 
and by birational transformations obtained via the iterated polynomial regularisation, one gets the Okamoto Hamiltonian~\eqref{OkHamP2}
\vspace*{-2ex} 
\begin{equation}\label{eq:P2_Oka}
H_{\text{II}}^{\text{Ok}}=
\begin{pmatrix}
    0 & -\frac{1}{2}\,z  & \frac{1}{2} \\
   -\alpha-\frac{1}{2} & 0 & 0 \\
   0 & -1 & 0
\end{pmatrix} \hspace{10ex} \includegraphics[width=.16\textwidth,valign=c]{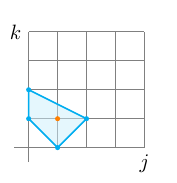} 
\end{equation} 
\vspace*{-2ex}

\noindent
characterised by a smaller area (2 rather than 3 as  in~\eqref{eq:P2_nonOka}). 
Note that in  \cite{MDGF} one more Hamiltonian for the second Painlev\'e equation was found 
\vspace*{-2ex}
\begin{equation}\label{eq:Ham_Bureau13}
H_{\text{II}}^{\text{Bu13}}\big(x(z),y(z);z\big)=x^2 y-2 \alpha  z\, x^2+z\, x^2+\alpha  z\, y-\frac{y^2}{4}\, \hspace{10ex} \includegraphics[width=.16\textwidth,valign=c]{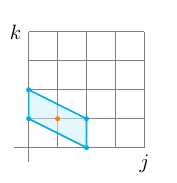} 
\end{equation}
\vspace*{-2ex}

\noindent
with the parallelogram-shaped Newton polygon of genus 1 and area equal to 2 (the labelling Bu13 refers to the system XIII in the classification of quadratic systems due to Bureau~\cite{MDGF}). The Hamiltonians~\eqref{eq:P2_nonOka},~\eqref{eq:P2_Oka} and~\eqref{eq:Ham_Bureau13} are pairwise birationally equivalent. We can choose either the Okamoto Hamiltonian~\eqref{eq:P2_Oka} or the Bureau Hamiltonian~\eqref{eq:Ham_Bureau13} as the fundamental one for $\text{P}_{\text{II}}$ for the minimal area of their polygons. Also, the two polygons share the same smallest highest total degree of~$3$.  

Lastly, let us consider the Okamoto Hamiltonians for other  Painlev\'e equations~\cite{okamoto1,okamoto2,okamoto4}, 
\begin{align} 
H_{\text{I}}^{\text{Ok}}\big(x(z),y(z);z\big) &= \frac{1}{2} y^2 - 2 x^3 - zx \,,\\[1ex]
H_{\text{III}}^{\text{Ok}}\big(x(z),y(z);z\big) &= \frac{1}{z} \left[ 2 y^2 x^2 - \left( 2 \eta_\infty\, z x^2 + (2\kappa_0 + 1) x - 2 \eta_0\, z \right)y + \eta_\infty (\kappa_0 + \kappa_\infty)zx \right] , \\[1ex]
H_{\text{V}}^{\text{Ok}}\big(x(z),y(z);z\big) &= \frac{1}{z} \left[ x(x-1)^2 y^2 - \left( \kappa_0 (x-1)^2 + \kappa_t\, x(x-1) - \eta\, zx \right) y + \kappa (x-1) \right] ,\\[1ex]
\begin{split} 
H_{\text{VI}}^{\text{Ok}}\big(x(z),y(z);z\big) &= \frac{1}{z(z-1)} \left[ x(x-1)(x-z) y^2 - \left[ \kappa_0\, (x-1)(x-z) + \kappa_1\, x(x-z) \right. \right. \\ & \left. \left. ~~+  (\kappa_t-1)\,x(x-1) \right] y + \kappa(x-t) \right],
\end{split} 
\end{align}
we observe that the areas of their Newton polygons is $2$ for the cases $\text{P}_{\text{I}}$ -- $\text{P}_{\text{IV}}$, and $7/2$ for both $\text{P}_{\text{V}}$ and $\text{P}_{\text{VI}}$:
\vspace*{-2ex}
\begin{equation}\label{eq:Okamoto_P1_P3_P6}
H^{\text{Ok}}_{\text{I}}\colon\includegraphics[width=.16\textwidth,valign=c]{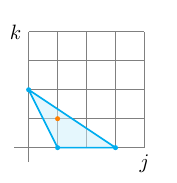} , \qquad  H^{\text{Ok}}_{\text{III}}\colon\includegraphics[width=.16\textwidth,valign=c]{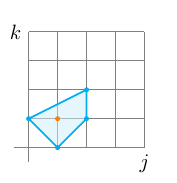} , \qquad H^{\text{Ok}}_{\text{V,VI}}\colon\includegraphics[width=.16\textwidth,valign=c]{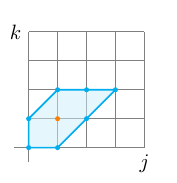}
\end{equation}
\vspace*{-2ex}

\noindent 
 We emphasise that without the constant term for the Okamoto Hamiltonians for $\text{P}_{\text{V}}$ and $\text{P}_{\text{VI}}$ the area of the corresponding Newton polygon would be $3$ instead of $7/2$. Note that all the Newton polygons associated with the Painlev\'e equations have genus $1$. At the moment, we do not know whether the areas of the Newton polygons of the Okamoto polynomial Hamiltonians can be made smaller by choosing some other Hamiltonians. However, we conjecture that the Newton polygons for the Okamoto polynomial Hamiltonians with rational coefficients have minimal areas for all Painlev\'e equations. 
 
We recall that the Hamiltonian functions associated with the Painlev\'e equations are not unique. For instance, a quartic Hamiltonian is given for the (modified) $\text{P}_{\text{V}}$ in~\cite[eq.~(5.62)]{MDTK2}, whose Newton polygon has area of $7/2$, and has a more symmetric shape than the one in~\eqref{eq:Okamoto_P1_P3_P6} and lower highest total degree ($4$ rather than $5$). Another example is in~\cite{GF1}, where one of the authors studied the cubic Hamiltonian associated with~$\text{P}_{\text{IV}}$. By means of the iterative polynomial regularisation (and the study of the Newton polygons) it was shown how the relation with~$\text{P}_{\text{IV}}$ can easily be obtained.

When dealing with problems of classification in the context of Painlev\'e and quasi-Painlev\'e equations, it is useful to identify a standard version of the system and to label it effectively (see the discussion in~\cite{MDTK2} and~\cite{GF}). As it was pointed out in~\cite{GF}, the method of iterative regularisation can be useful for this purpose, leading to find ``simpler'' looking equivalent  systems. Also, the study of different shapes of Newton polygons arising at intermediate steps in the iterative regularisation is useful for solving the so-called (quasi)-Painlev\'e equivalence problem. The equivalence problem can be highly relevant in applications. For instance, in the context of orthogonal polynomials this might be useful to unveil some relations with Painlev\'e equations~\cite{VAbook}. In this context, the iterative regularisation has already been successfully used in the case of Charlier orthogonal polynomials~\cite{GFCR} and Krawtchouk orthogonal polynomials~\cite{GFCR2}, and might be of interest in several related problems of random matrix models (see e.g. for $\text{P}_{\text{IV}}$~\cite{Chen1,Chen2,Chen3}). 

Here, we refine the labelling of (quasi)-Painlev\'e equations, originally proposed in \cite{GF}, and use the following notation  
\begin{equation}\label{eq:labelling}
((k_1)_{\ell_1},\dots,(k_m)_{\ell_m}, k_{m+1},\ldots, k_n)\,, 
\end{equation}
meaning that we have $n$ final Hamiltonian systems, whose solutions have Puiseux expansions~\eqref{eq:puiseux} with branching orders $1/{k_1}, \dots, 1/{k_n}$. If $\ell$ in~\eqref{eq:puiseux} is such that $|\ell|=\ell_i>1$  for $i=1, \dots, m$, we might need to introduce the so-called $\ell_{i}$-fold covering transformation (see appendix~\ref{app:regularisation}).   
 For instance, the solutions of $\text{P}_{\text{I}}$ have  second order movable poles and the corresponding Hamiltonian system is regularised by using a 2-fold covering, see \cite{IwasakiOkada}. Therefore, the notation $((1)_2)$ can be used. Instead, the system associated with $\text{P}_{\text{II}}$ is such that there are two types of first order movable poles, and the labelling $(1,1)$ can be used.
In the regularisation process of the Hamiltonian system for $\text{P}_{\text{IV}}$ (see, for instance, \cite{Takano97}) there are tree cascades, therefore we can use notation (1,1,1). When the covering transformation is not used in the process, we speak of polynomial or rational regularisation. For instance, $\text{P}_{\text{I}}$ can be rationally regularised and one can just use the notation (1). Therefore, when labelling quasi-Painlev\'e equations using the method above we should distinguish between rational or polynomial regularisation.

In the following, section~\ref{sec:genP4} and~\ref{sec:modgenP4} are devoted to the study of two different Hamiltonians for equations of type $(1,1,1)$ via the iterative regularisation. Section~\ref{sec:quasiP4} is dedicated to the analysis of the mixed case $(2,2,1)$, and finally section~\ref{sec:open} contains some discussion and open questions.

\section{\texorpdfstring{Generalised Okamoto Hamiltonian for $\text{P}_{\text{IV}}$}{gen P4}}
\label{sec:genP4}

To be self-contained we briefly describe the polynomial regularisation in appendix~\ref{app:regularisation}. 
In this section, we consider the following Hamiltonian
\begin{equation}\label{eq:Ok1}
   H^{\text{Ok}}_{\text{gen}}(x(z),y(z);z) = \beta_0(z)\, x- \left(\beta_1(z)+\beta_2(z)\, x+x^2\right)y+2 x y^2\,,
\end{equation}
with coefficient matrix and Newton polygon as in \eqref{eq:HamOkamoto1}
\vspace*{-2ex}
\begin{equation}\label{eq:Ok1_NP_coeffmat}
    H^{\text{Ok}}_{\text{gen}}=\begin{pmatrix}
 0 & -{\beta_1(z)} & 0 \\
 \beta_0(z) & -{\beta_2}(z) & 2 \\
 0 & -1 & 0 \\
  \end{pmatrix} \hspace{10ex} \includegraphics[width=.16\textwidth,valign=c]{Okamoto_1.pdf} 
\end{equation}
\vspace*{-2ex}

\noindent
where the coefficients $\beta_i(z)$, for $i=0\,,1,2$, are unknown analytic functions of the independent variable. 
This is a generalised version of the Okamoto Hamiltonian~\eqref{OkHamP4}, and we show here how to relate the two. Since we shall omit some details in the next sections, this example will be useful for illustrative purposes. We shall pay attention to the Newton polygons for the polynomial Hamiltonians in the final chart systems, focusing on their shape, area, genus and highest total degree. As mentioned above, we conjectured that Okamoto Hamiltonians are the fundamental ones in the set of Hamiltonians which are (at most) rational in all variables. 

We start the regularisation process $r_0$ by looking for points of indeterminacy in $\mathbb{P}^1 \times \mathbb{P}^1$, identifying three points: 
\vspace{1ex}
\begin{equation}\label{eq:basepoints_1}
\hspace*{-2ex} 
\includegraphics[width=.25\textwidth,valign=c]{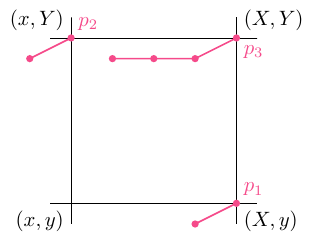}~~
\begin{aligned} X &= \frac{1}{x} \,, \quad Y = \frac{1}{y}\,, \\[1ex] 
p_1 &\colon (X,y) = (0\,,0), \quad p_2 \colon (x,Y) = (0\,,0), \quad p_3 \colon (X,Y) = (0\,,0).
\end{aligned} 
\end{equation}
\vspace{1ex}
\noindent
In the following, we list the points in the different cascades originating at $p_1, p_2, p_3$, the resonance conditions for the system to be of Painlev\'e type, 
and present the expressions of the polynomial Hamiltonians in the final charts with the related Newton polygons.

The cascade originating at $p_1$ in \eqref{eq:basepoints_1} for the initial regularisation $r_0$ is 
    \begin{equation}
        p_1 \colon (X,y) = (0\,,0)~\leftarrow~ (u_1^{(1)},v_1^{(1)}) = (0\,,\beta_0(z)) \,,
    \end{equation}
    with condition for the system to be regular (without any points of indeterminacy in the last chart with coordinates $(u_2^{(1)},v_2^{(1)})$) being
    \begin{equation}
        \beta_0'(z) = 0 \, \implies  \beta_0(z) = \beta_0 \,, 
    \end{equation}
    i.e.\ $\beta_0$ is a complex constant. 
    Including this condition, the Hamiltonian in the variables of the final chart of this cascade is given by (omitting the upper index $(1)$ in the coordinates $(u_2^{(1)},v_2^{(1)})$)
    \begin{equation}
    \begin{split}
     H^{(1)}_{\text{gen},0}\big(u_2(z),v_2(z);z\big) &= 
        \left(\beta _1(z)-4 \beta _0\right)u_2^2 v_2 +\left(v_2 \beta _2(z)+\beta _0 \beta _1(z)-2 \beta _0^2\right)u_2 -2
   u_2^3 v_2^2+v_2\,, 
    \end{split} 
    \end{equation}
    with coefficient matrix and Newton polygon with area equal to 2 as
     \begin{equation}
H^{(1)}_{\text{gen},0}=\begin{pmatrix}
 0 & 1 & 0 \\
\beta_0 ( \beta_1(z) -2  \beta_0) &  \beta_2(z) & 0 \\
 0 & \beta_1(z)-4  \beta_0  & 0 \\
 0 & 0 & -2 \\
  \end{pmatrix} \hspace{10ex} \includegraphics[width=.16\textwidth,valign=c]{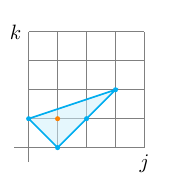} 
\end{equation}
    \vspace*{-2ex}

\noindent
The lower index $0$ in $H_{\text{gen},0}^{(1)}$ indicates the regularisation $r_0$. 

The cascade originating at $p_2$ in \eqref{eq:basepoints_1} in the regularisation $r_0$ is given by
\begin{equation}
 \begin{split} 
        p_2 &\colon (x,Y) = (0\,,0)~\leftarrow~ (\widetilde{u}_1^{(2)},\widetilde{v}_1^{(2)}) = \left(\tfrac{1}{2}\beta_1(z)\,,0\right)\,,
    \end{split} 
    \end{equation}
where we used the intermediate change of variables $v_1^{(2)}\to 1/v_1^{(2)}$ leaving $u_1^{(2)}$ unchanged (see appendix~\ref{subapp:intermediate}). 
The resulting final chart system in $(u_2,v_2)$ is regular, provided that
\begin{equation}\label{eq:res_cond_beta1}
        \beta_1'(z) = 0 \, \implies  \beta_1(z) = \beta_1\, ,       
\end{equation}
hence $\beta_1$ is a constant as well. Additionally, the final chart system is Hamiltonian with the polynomial Hamiltonian represented by the following coefficient matrix and the diamond-shaped Newton polygon with area equal to $2$
 \vspace*{-2ex}
 \begin{equation}\label{eq:H02}
H^{(2)}_{\text{gen},0}=\left(
\begin{array}{ccc}
 0 & \beta _1 & 0 \\
 \beta _0-\frac{1}{2} \beta _1 & -\beta _2(z) & 2 \\
 0 & -1 & 0 \\
\end{array}
\right) \hspace{10ex} \includegraphics[width=.16\textwidth,valign=c]{Okamoto_1.pdf} 
\end{equation}
\vspace*{-2ex}

\noindent
Note that without using the intermediate change of coordinates in the cascade, the regularisation would have been along the cascade
 \begin{equation}
 \begin{split} 
        p_2 &\colon (x,Y) = (0\,,0)~\leftarrow~ (U_1^{(2)},V_1^{(2)}) = (\tfrac{1}{2}\beta_1(z),0)\,,
    \end{split} 
    \end{equation}
    with the same condition for the final chart system to be regular. However, the polynomial Hamiltonian of the final chart system in $\big(U_2^{(2)}, V_2^{(2)}\big)$ has 
    coefficient matrix and Newton polygon with area $2$ given by 
    \vspace*{-2ex}
 \begin{equation}
H^{(2)}_{\text{gen},\text{alt}}=\left(
\begin{array}{cccc}
 0 & \frac{1}{4}\beta _1(\beta _1-2 \beta _0 ) & 0 & 0 \\
 -2 & \beta _2(z) & \beta _1-\beta _0 & 0 \\
 0 & 0 & 0 & 1 \\
\end{array}
\right) \hspace{10ex} \includegraphics[width=.16\textwidth,valign=c]{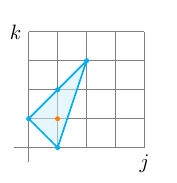} 
\end{equation}
\vspace*{-2ex}

\noindent 
hence presenting more complicated coefficients in comparison with~\eqref{eq:H02}. 

Lastly, the cascade originating at $p_3$ in \eqref{eq:basepoints_1} in the regularisation $r_0$ is given by
 \begin{equation}
 \begin{split} 
        p_3 &\colon (X,Y) = (0\,,0)~\leftarrow~ (\widetilde{u}_1^{(3)},\widetilde{v}_1^{(3)}) = (0\,,\tfrac{1}{2}) ~\leftarrow~ (u_2^{(3)},v_2^{(3)}) = (0\,,\tfrac{1}{2}\beta_2(z)) ~\leftarrow~\\[1ex] 
        &~\leftarrow~ (u_3^{(3)},v_3^{(3)}) =  (0\,, \tfrac{1}{2}(\beta_1-2\beta_0-\beta_2'(z))  ) \,.
    \end{split} 
    \end{equation}
    The condition for the system to be regular and polynomial (without any points of indeterminacy in the last chart $(u_4^{(3)},v_4^{(3)})$) is 
    \begin{equation}
        \beta_2''(z) = 0 \,\implies \beta_2(z) = a_1 + a_2\, z\,, 
    \end{equation}
    with $a_1,a_2$ two complex constants. 
    The  coefficient matrix and the area $2$ Newton polygon are
    \vspace*{-2ex}
    \begin{equation}\label{eq:H03_gen}
H^{(3)}_{\text{gen},0}=\left(
\begin{array}{ccc}
 0 & -1 & 0 \\
 -\frac{1}{2} \left(a_2+2 \beta _0\right) \left(a_2+2 \beta _0-\beta _1\right) & -a_2 z-a_1 & 0 \\
 0 & 2 a_2+4 \beta _0-\beta _1 & 0 \\
 0 & 0 & -2 \\
\end{array}
\right)\hspace{5ex} \includegraphics[width=.16\textwidth,valign=c]{r0_H1.pdf} 
\end{equation}
\vspace*{-2ex}

Overall, the regularisation $r_0$ of the generalised Okamoto Hamiltonian for $\text{P}_{\text{IV}}$ in terms of the Newton polygons is summarised as 
\vspace*{-2ex}
\begin{equation}
    \includegraphics[width=.75\textwidth,valign=c]{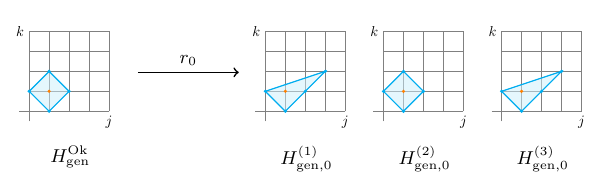}
\end{equation}
\vspace*{-2ex}

The resonance conditions for all the systems in the final charts (emerging after the regularisation process $r_0$ applied to~\eqref{eq:Ok1}) to be polynomial yield the following coefficient matrix
\begin{equation}\label{compare}
H^{\text{Ok}}_{\text{gen}}= \begin{pmatrix}
 0 & -{\beta_1} & 0 \\
 \beta_0 & -a_1-{a_2}\,z & 2 \\
 0 & -1 & 0 \\
  \end{pmatrix}
\end{equation}
with constants $a_1,a_2,\beta_0\,, \beta_1 \in \mathbb{C}$. The corresponding Hamiltonian system gives the fourth Painlev\'e equation after the matching
\begin{equation}
    \beta_0 = \vartheta_{\infty}\,, \qquad \beta_1=2\,\kappa_0\,, \qquad a_1=0\,, \qquad a_2=2\,,
\end{equation}
from the comparison of \eqref{compare} with \eqref{eq:HamOkamoto1}.

\section{\texorpdfstring{Modified generalised Okamoto Hamiltonian for $\text{P}_{\text{IV}}$}{mod gen P4}}
\label{sec:modgenP4}

We now consider a modified version of the generalised Okamoto Hamiltonian \eqref{eq:Ok1} 
\begin{equation}\label{eq:Ok2}
   H^{\text{Ok}}_{\text{mod}}(x(z),y(z);z) = \beta (z) x^2+\beta_0(z) x- \left(\beta_1(z)+\beta_2(z) x+x^2\right)y+2 x y^2\,, 
\end{equation}
with the additional term with coefficient function $\beta(z)$. The Newton polygon has area $5/2$:
\vspace*{-2ex}
\begin{equation}\label{eq:H_Ok_mod}
H^{\text{Ok}}_{\text{mod}}=\left(
\begin{array}{ccc}
 0 & -\beta _1(z) & 0 \\
 \beta _0(z) & -\beta _2(z) & 2 \\
 \beta (z) & -1 & 0 \\
\end{array}
\right) \hspace{10ex} \includegraphics[width=.16\textwidth,valign=c]{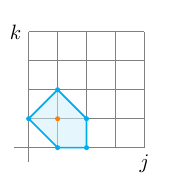}     
\end{equation}
\vspace*{-2ex}

\noindent

The second order differential equation for $x=x(z)$ is given by
\begin{equation}\label{equation for x}
\begin{split} 
x''&=\frac{(x')^2}{2x}+\frac{3}{2}x^3 +(2\beta_2(z)-8\beta(z))x^2 +\left(\beta_1(z)-4\beta_0(z)+\frac{\beta_2(z)^2}{2}-\beta_2'(z)\right)x\\
&~~~~-\beta_1'(z)-\frac{\beta_1(z)^2}{2x}\,.
\end{split} 
\end{equation}
We shall show that we can obtain the diamond-shaped Hamiltonian in the second iteration of the polynomial regularisation. We will see that we obtain Newton polygons with area $2$ already in the first regularisation $r_0$, with triangular shapes. However, these are not minimal in the sense that they correspond to a larger highest total degree in the Hamiltonian than that of the diamond shape (i.e.\ $5$ rather than $3$). 

The polynomial regularisation $r_0$ is obtained via the following cascades of blow-ups, from the  points of indeterminacy in $\mathbb{P}^1 \times \mathbb{P}^1$:
\vspace{1ex}
\begin{equation}\label{eq:basepoints_2}
\hspace*{-2ex} 
\includegraphics[width=.25\textwidth,valign=c]{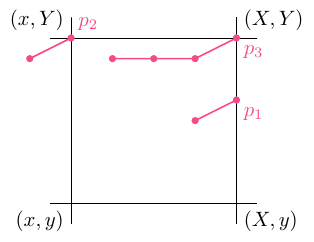}
\begin{aligned} X &= \frac{1}{x} \,, \quad Y = \frac{1}{y}\,, \\[1ex] 
p_1 &\colon (X,y) = (0\,,\beta(z)), ~~ p_2 \colon (x,Y) = (0\,,0), ~~ p_3 \colon (X,Y) = (0\,,0).
\end{aligned} 
\end{equation}
\vspace{1ex}

\noindent
The cascade originating at $p_1$ is
    \begin{equation}
        p_1 \colon (X,y) = (0\,,\beta(z))~ \leftarrow ~ (u_1^{(1)}, v_1^{(1)}) = (0\,,2 \beta(z)^2+\beta'(z)-\beta(z) \beta_2(z)+\beta_0(z))\,,
    \end{equation}
    and the resonance condition for the system to be regular in the final chart is equivalent to
    \begin{equation}
 \beta_0(z)=\beta(z)\beta_2(z)-2\beta(z)^2-\beta'(z)+b_0\,, 
    \end{equation}
with $b_0$ integration constant. The system in the final chart is Hamiltonian with the coefficient matrix and Newton polygon with area $2$
 \vspace*{-1ex}
\begin{equation}\label{eq:Okmod_H0_1}
H^{(1)}_0=\left(
\begin{array}{ccc}
 0 & 1 & 0 \\
 b_0 (\beta _1(z)-2 b_0) & \beta _2(z)-4 \beta (z) & 0 \\
 0 & \beta _1(z)-4 b_0 & 0 \\
 0 & 0 & -2 \\
\end{array}
\right)\hspace{10ex} \includegraphics[width=.16\textwidth,valign=c]{r0_H1.pdf} 
\end{equation}
\vspace*{-2ex}

\noindent
The second cascade is 
 \begin{equation}\label{eq:Ok_2_cascade_1}
        p_2 \colon (x,Y) = (0\,,0)~ \leftarrow ~ (\widetilde{u}_1^{(2)}, \widetilde{v}_1^{(2)}) = (0\,,\tfrac{1}{2}\beta_1(z))\,, 
    \end{equation}
    with the intermediate change of variables $(u_1^{(2)},v_1^{(2)}) \mapsto (\widetilde{u}_1^{(2)}=u_1^{(2)}, \widetilde{v}_1^{(2)} = 1/v_1^{(2)})$. The resonance condition is analogous to~\eqref{eq:res_cond_beta1} $\beta_1'(z)=0$, i.e.\ $\beta_1(z) = \beta_1$ constant. The final chart Hamiltonian for this cascade is represented by the coefficient matrix
     \vspace*{-1ex}
    \begin{equation}\label{eq:Okmod_H0_2}
    H^{(2)}_0=
    \left(
\begin{array}{ccc}
 0 & \beta _1 & 0 \\
 b_0-\beta(z)(2 \beta (z)-\beta _2(z))-\frac{1}{2}\beta _1-\beta '(z) & -\beta _2(z)  & 2 \\
 \beta (z) & -1 & 0 \\
\end{array}
\right)\hspace{5ex} \includegraphics[width=.16\textwidth,valign=c]{r0_mod_H0.pdf} 
\end{equation}
    \vspace*{-2ex}

    \noindent
   We see that the Newton polygon is still pentagon-shaped, as the initial one.

As in the previous case, we could have blown up the point $(U_1^{(2)},V_1^{(2)})=(\tfrac{1}{2}\beta_1(z),0)$ without any intermediate change of variables. This would have given us a polynomial Hamiltonian in the final chart, provided that the same resonance condition is satisfied ($\beta_1$ constant):
\vspace*{-2ex}
\begin{equation}
H^{(2)}_{\text{alt}}=\left(
\begin{array}{ccccc}
 0 & \frac{1}{4}\beta _1(\beta _1-2 \beta _0(z)) & -\frac{1}{4} \beta _1^2 \,\beta
   (z) & 0 & 0 \\
 -2 & \beta _2(z) & \beta _1-\beta _0(z) & -\beta _1 \beta (z) & 0 \\
 0 & 0 & 0 & 1 & -\beta (z) \\
\end{array}
\right)
\hspace{3ex} \includegraphics[width=.16\textwidth,valign=c]{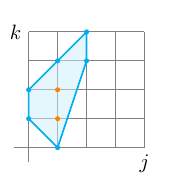} 
\end{equation}
    \vspace*{-2ex}

    \noindent
 with area equal to 4. As it is evident from the picture, the genus of the algebraic curve describing the Hamiltonian is increased by one. 
We do not know whether we could obtain an even higher genus in some step of the iterative regularisation process. 

Imposing the resonance conditions obtained  from the first two cascades, the cascade originating at $p_3$ is given by 
    \begin{equation}
    \begin{split} 
        p_3\colon (X,Y)&=(0\,,0)~ \leftarrow (\widetilde{u}_1^{(3)}, \widetilde{v}_1^{(3)}) = (0\,,\tfrac{1}{2}) ~ \leftarrow ~ (u_2^{(3)}, v_2^{(3)}) = (0\,,\tfrac{1}{2}(\beta_2(z)-2\beta(z)))~ \leftarrow \\[1ex] ~ &\leftarrow ~ (u_3^{(3)}, v_3^{(3)}) = (0\,,\tfrac{1}{2}(\beta_1(z)+4\beta'(z)-\beta_2'(z)-2b_0)),
        \end{split} 
    \end{equation}
alongside with the resonance condition 
\begin{equation}
    \beta_2''(z)=4\beta''(z)\, \implies \beta_2(z)=4\beta(z)+a_1+a_2 z\,, 
\end{equation} 
with $a_1, a_2$ constants.  The coefficient matrix of the Hamiltonian of the final chart system in the chart $(u_4^{(3)},v_4^{(3)})$ is given by
\begin{equation}\label{eq:Okmod_H0_3}
\hspace*{-1ex} H^{(3)}_0=\left(
\begin{array}{ccc}
 0 & -1 & 0 \\
 \frac{1}{2} \left(a_2+2 b_0\right) \left(\beta_1-a_2-2 b_0\right) & -(a_1+ a_2 z) & 0 \\
 0 & 2 a_2-\beta _1+4 b_0  & 0 \\
 0 & 0 & -2 \\
\end{array}
\right) \hspace{3ex} \includegraphics[width=.16\textwidth,valign=c]{r0_H1.pdf} 
\end{equation}
    \vspace*{-2ex}

    \noindent
with the Newton polygon triangular-shaped as in the first cascade and with area 2. As we can see $H_0^{(3)}$ does not depend on $\beta(z)$ any more, and it coincides with the Hamiltonian $H^{(3)}_{\text{gen},0}$ in~\eqref{eq:H03_gen} obtained at the end of the regularisation $r_0$ in the generalised case.

Although we could already reduce the ODE \eqref{equation for x} to $\text{P}_{\text{IV}}$ via a simple re-scaling, $\beta$ is still present in the Hamiltonian 
\begin{equation}
    H^{\text{Ok}}_{\text{mod}}=\left(
\begin{array}{ccc}
 0 & -\beta _1 & 0 \\
 2 \beta (z)^2+\beta (z) \left(a_1+a_2 z\right)-\beta '(z)+b_0 & -(a_1+a_2 z+4
   \beta (z)) & 2 \\
 \beta (z) & -1 & 0 \\
\end{array}
\right) 
\end{equation} 
and the Newton polygon associated with it is not diamond-shaped, but still pentagonal. After imposing the resonance conditions at the end of each cascade, the coefficient function $\beta(z)$ is still unknown and appears in~$H^{(2)}_0$ in~\eqref{eq:Okmod_H0_2} only.

To summarise, the action of the regularisation $r_0$ in terms of Newton polygons is:
\vspace*{-2ex}
\begin{equation}
    \includegraphics[width=.75\textwidth,valign=c]{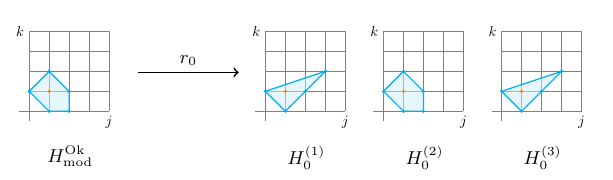}
\end{equation}
\vspace*{-2ex}

\noindent
Now we investigate what shapes of Newton polygons we obtain after the next step of the iterative regularisation procedure $r_1$. 

Let us consider the final chart system associated with the Hamiltonian $H^{(1)}_0$ in~\eqref{eq:Okmod_H0_1}, including all the resonance conditions obtained after $r_0$, as our original system. We introduce the compactification $\mathbb{P}^1 \times \mathbb{P}^1$ and study the Hamiltonian system associated with $H_0^{(1)}$, looking for the base points:
\begin{equation}\label{eq:basepoints_3}
\hspace*{-2ex} 
\includegraphics[width=.25\textwidth,valign=c]{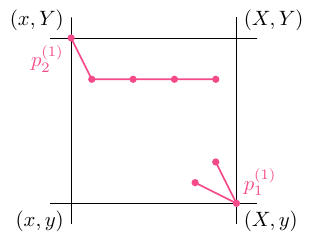} \hspace{5ex}
\begin{aligned} X &= \frac{1}{x} \,, \quad Y = \frac{1}{y}\,, \\[1ex] 
p_1^{(1)} &\colon (X,y) = (0\,,0), ~~ p_2^{(1)} \colon (x,Y) = (0\,,0).
\end{aligned} 
\end{equation}
The regularisation $r_1$ of the system $H^{(1)}_0$ gives rise to three final chart systems, since the cascade originating from $p_1^{(1)}$ in~\eqref{eq:basepoints_3} splits into two sub-cascades: 
\begin{equation}\label{eq:H01_p11}
    \begin{tikzpicture}[baseline={(0, .1cm-\MathAxis pt)}]
        \path (0,0) coordinate (a) -- ++(4.25,.5) coordinate (b); 
        \path (a) -- ++(4.6,-.5) coordinate (c); 
        \node (a1) at (a) {$p_1^{(1)}\colon (X,y) = (0\,,0)$} node[right=1.5cm of a] (c1) {};
        \node (a2) at (b) {$(u_1^{(1,-)},v_1^{(1,-)}) = (0\,,-b_0)\,,$} node[left=1.8cm of b] (c2) {};
        \node (a3) at (c) {$(u_1^{(1,+)},v_1^{(1,+)}) = (0\,,\tfrac{1}{2}\beta_1-b_0)\,.$} node[left=2.2cm of c] (c3) {};
        \draw[<-] (c1) -- (c2);
        \draw[<-] (c1) -- (c3);
    \end{tikzpicture}
\end{equation}
The Hamiltonians and their associated Newton polygons for the final charts $(u_2^{(1,\mp)},v_2^{(1,\mp)})$ are:
\vspace*{-2ex}
\begin{align}
\label{eq:H1minus}
    H^{(1,-)}_1&=\left(
\begin{array}{ccc}
 0 & -\beta_1 & 0 \\
 b_0 & -(a_1+a_2 z) & 2 \\
 0 & -1 & 0 \\
\end{array}
\right) \hspace{10ex} \includegraphics[width=.16\textwidth,valign=c]{Okamoto_1.pdf} \\[-2ex]
\label{eq:H1plus}
H^{(1,+)}_1&=
\left(
\begin{array}{ccc}
 0 & \beta _1 & 0 \\
b_0-\frac{1}{2}\beta_1    & -(a_1+a_2 z) & 2 \\
 0 & -1 & 0 \\
\end{array}
\right) \hspace{10ex} \includegraphics[width=.16\textwidth,valign=c]{Okamoto_1.pdf} 
\end{align}
\vspace*{-3ex}

\newpage
The cascade originating in $p_2^{(1)}$ in \eqref{eq:basepoints_3} is longer and requires an intermediate change of coordinates:
\begin{equation}\label{eq:H01_p21}
    \begin{split}
        p_2^{(1)} &\colon (x,Y)=(0\,,0) \,\, \leftarrow \,\, (u_1^{(1,2)},v_1^{(1,2)})=(0\,,0) \,\, \leftarrow \,\, (u_2^{(1,2)},v_2^{(1,2)})=(0\,,0) \,\, \leftarrow\\[1ex]
        &\leftarrow \,\, (\widetilde{u}_3^{(1,2)},\widetilde{v}_3^{(1,2)})=(0\,,\tfrac{1}{2}) \,\, \leftarrow \,\, (u_4^{(1,2)},v_4^{(1,2)})=(0\,,\tfrac{1}{2}(a_1+a_2\,z)) \,\, \leftarrow \\[1ex]
        &\leftarrow \,\, (u_5^{(1,2)},v_5^{(1,2)})=(0\,,\tfrac{1}{2}(\beta_1-a_2-4b_0))\,. 
    \end{split}
\end{equation}
The final system in the chart $(u_6^{(1,2)},v_6^{(1,2)})$ is described by the following Hamiltonian and related Newton polygon 
\vspace*{-2ex}
\begin{equation}\label{eq:H12}
{H^{(1,2)}_1}=
\left(
\begin{array}{ccc}
 0 & -1 & 0 \\
 b_0(\beta_1-2b_0)  & a_1+a_2 z & 0 \\
 0 & \beta_1-4b_0 & 0 \\
 0 & 0 & -2 \\
\end{array}
\right) \equiv H_0^{(1)} \hspace{10ex} \includegraphics[width=.16\textwidth,valign=c]{r0_H1.pdf} 
\end{equation}

\noindent
coinciding with the initial Hamiltonian system $H_0^{(1)}$. 
Therefore, the scheme for the regularisation~$r_1$ of $H_0^{(1)}$ is the following 
\vspace*{-2ex}
\begin{equation}\label{eq:r1_H01_scheme}
    \includegraphics[width=.8\textwidth,valign=c]{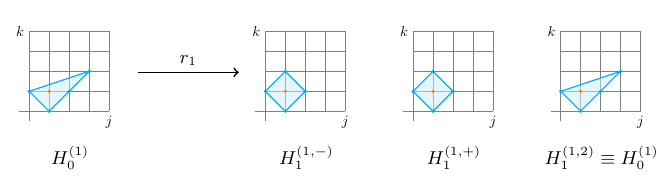}
\end{equation}
\vspace*{-2ex}

Next, we consider the system described by the Hamiltonian $H^{(2)}_0$ in~\eqref{eq:Okmod_H0_2} (with the resonance conditions) to be the input of the regularisation $r_1$. We proceed with the compactification of the phase space $\mathbb{P}^1 \times \mathbb{P}^1$, in which we identify the base points. We find that the configuration of the cascades is the same as in~\eqref{eq:basepoints_2}. The sequences of blow-ups are: 
\begin{align}
    p_1^{(2)} &\colon (X,y) = (0\,,\beta(z))~ \leftarrow ~ (u_1^{(2,1)}, v_1^{(2,1)}) = (0\,,2 \beta '(z)-\tfrac{1}{2}\beta_1+ b_0)\,, \\[1ex]
    p_2^{(2)} &\colon (x,Y) = (0\,,0) ~ \leftarrow ~ (u_1^{(2,2)}, v_1^{(2,2)}) = (0\,,-\tfrac{1}{2}\beta_1)\,, \\[1ex]
    \begin{split}
    p_3^{(2)} &\colon (X,Y) = (0\,,0) ~\leftarrow~ (\widetilde{u}_1^{(2,3)},\widetilde{v}_1^{(2,3)}) = (0\,,\tfrac{1}{2}) ~\leftarrow~\\[1ex] 
        &~\leftarrow~ (u_2^{(2,3)},v_2^{(2,3)}) = (0\,,\beta(z)+\tfrac{1}{2}(a_1+a_2\,z)) ~\leftarrow~ \\[1ex] 
        &~\leftarrow~  (u_3^{(2,3)},v_3^{(2,3)}) =  (0\,, -\tfrac{1}{2}{a_2}-b_0-2\beta'(z) ) \,.
    \end{split} 
\end{align}
The condition for the regularisation of the system after $r_1$ is 
\begin{equation}
    \beta''(z)=0 ~\implies~ \beta(z) = b_1+b_2\,z\,. 
\end{equation}
\vspace*{-4ex}

\noindent
The coefficients matrices for the final Hamiltonians are, respectively, 
\begin{equation}
\label{eq:H21}
\hspace*{-2ex}H^{(2,1)}_1=
\left(
\begin{array}{ccc}
 0 & 1 & 0 \\
 (\beta_1-2 b_0)b_0  & a_1+a_2 z & 0 \\
 0 & \beta_1-4b_0 & 0 \\
 0 & 0 & -2 \\
\end{array}
\right) \equiv H_0^{(1)} 
\end{equation}
\begin{align}  
\label{eq:H22}
\hspace*{-2ex}H^{(2,2)}_1&=
\left(
\begin{array}{ccc}
 0 & -\beta_1 & 0 \\
b_0-b_2 + (b_1 + b_2 z) (a_1 + 2 b_1 + (a_2 + 2 b_2) z)  & -(a_1+4b_1+(a_2+4b_2) z) & 2 \\
 b_1 + b_2 z & -1 & 0 \\
\end{array}
\right) \\[1ex]
\label{eq:H23}
\hspace*{-2ex}H^{(2,3)}_1&=
\left(
\begin{array}{ccc}
 0 & -1 & 0 \\
  (\frac{1}{2}a_2+ b_0) (\beta_1-a_2+2 b_0) & -(a_1+a_2 z) & 0 \\
 0 & 2a_2-\beta_1+4b_0 & 0 \\
 0 & 0 & -2 \\
\end{array}
\right) \equiv H_0^{(3)}\,,  
\end{align}
where the coefficient matrix for $H_{1}^{(2,1)}$ coincides with $H_0^{(1)}$ in~\eqref{eq:Okmod_H0_1} with the resonance conditions, and the coefficient matrix for $H_{1}^{(2,3)}$ coincide with that of $H_0^{(3)}$ in~\eqref{eq:Okmod_H0_3}.  
The iterative step regularisation $r_1$ from $H_0^{(2)}$ in~\eqref{eq:Okmod_H0_2} is
\vspace*{-2ex}
\begin{equation}
    \includegraphics[width=.8\textwidth,valign=c]{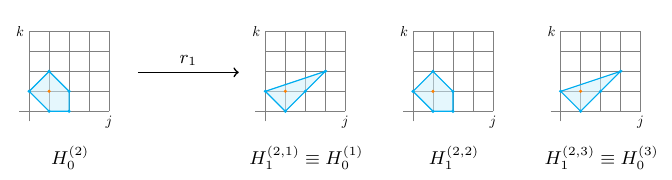}
\end{equation}
\vspace*{-2ex}

Lastly, for the regularisation $r_1$ applied to $H^{(3)}_0$ in~\eqref{eq:Okmod_H0_3}, the result is analogous to the one obtained for $H^{(1)}_0$ and schematised in~\eqref{eq:r1_H01_scheme} in terms of shapes of Newton polygons, with different Hamiltonians for the final systems: 
\vspace*{-2ex}
\begin{equation}
    \includegraphics[width=.8\textwidth,valign=c]{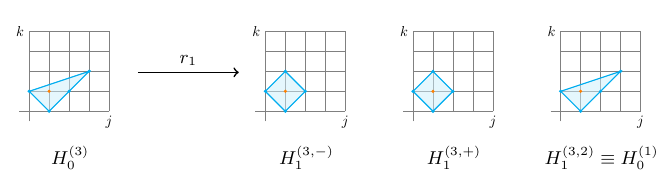}
\end{equation}
\vspace*{-2ex}

\noindent
The location of base points $p_1^{(3)}, p_2^{(3)}$ coincides with that of the points $p_1^{(1)}, p_2^{(1)}$ in~\eqref{eq:basepoints_3}, and the cascades obtained in the regularisation $r_1$ are analogous of those reported in~\eqref{eq:H01_p11} and~\eqref{eq:H01_p21}:
\begin{align}
 &\begin{tikzpicture}[baseline={(0, .1cm-\MathAxis pt)}]
        \path (0,0) coordinate (a) -- ++(5,.5) coordinate (b) -- ++(.4,-1) coordinate (c);
        \node (a1) at (a) {$p_1^{(3)}\colon (X,y) = (0\,,0)$} node[right=1.5cm of a] (c1) {};
        \node (a2) at (b) {$(u_1^{(3,-)},v_1^{(3,-)}) = (0\,,\tfrac{1}{2}a_2+2b_0)\,,$} node[left=2.3cm of b] (c2) {};
        \node (a3) at (c) {$(u_1^{(3,+)},v_1^{(3,+)}) = (0\,,b_0-\tfrac{1}{2}a_2-\tfrac{1}{2}\beta_1)\,,$} node[left=2.75cm of c] (c3) {};
        \draw[<-] (c1) -- (c2);
        \draw[<-] (c1) -- (c3);
    \end{tikzpicture} \\[2ex]
\begin{split} 
&~p_2^{(3)} \colon (x,Y)=(0\,,0) \,\, \leftarrow \,\, (u_1^{(3,2)},v_1^{(3,2)})=(0\,,0) \,\, \leftarrow \,\, (u_2^{(3,2)},v_2^{(3,2)})=(0\,,0) \,\, \leftarrow\\[1ex]
        &\hspace{4ex} \leftarrow \,\, (\widetilde{u}_3^{(3,2)},\widetilde{v}_3^{(3,2)})=(0\,,\tfrac{1}{2}) \,\, \leftarrow \,\, (u_4^{(3,2)},v_4^{(3,2)})=(0\,,\tfrac{1}{2}(a_1+a_2\,z)) \,\, \leftarrow \\[1ex]
        &\hspace{4ex} \leftarrow \,\, (u_5^{(3,2)},v_5^{(3,2)})=(0\,,\tfrac{1}{2}(\beta_1-a_2-4b_0))\,. 
\end{split} 
\end{align}

\noindent
The coefficients matrices for the Hamiltonians in the final charts $(u_2^{(3,-)},v_2^{(3,-)})$, $(u_2^{(3,+)},v_2^{(3,+)})$ and $(u_6^{(3,2)},v_6^{(3,2)})$ respectively are: 
\vspace*{-2ex}
\begin{align}
\label{eq:H3minus}
H^{(3,-)}_1&=
\left(
\begin{array}{ccc}
 0 & \beta_{1} & 0 \\
 \frac{1}{2}a_{2}+b_0 & a_{1}+a_{2} z & 2 \\
 0 & 1 & 0 \\
\end{array}
\right) \hspace{10ex} \includegraphics[width=.16\textwidth,valign=c]{Okamoto_1.pdf} \end{align}
\begin{align} 
\label{eq:H3plus}
H^{(3,+)}_1&=
\left(
\begin{array}{ccc}
 0 & -\beta_{1} & 0 \\
 \frac{1}{2}a_{2}+b_0-\frac{1}{2}\beta_1 & a_{1}+a_{2} z& 2 \\
 0 & 1 & 0 \\
\end{array}
\right) \hspace{10ex} \includegraphics[width=.16\textwidth,valign=c]{Okamoto_1.pdf} \\[-1ex]
\label{eq:H32}
{H^{(3,2)}_1}&=
\left(
\begin{array}{ccc}
 0 & 1 & 0 \\
 (\beta_1-2 b_0)b_0  & a_1+a_2 z & 0 \\
 0 & \beta_1-4b_0 & 0 \\
 0 & 0 & -2 \\
\end{array}
\right) \equiv H_0^{(1)}  \hspace{10ex} \includegraphics[width=.16\textwidth,valign=c]{r0_H1.pdf} 
\end{align}
\vspace*{-2ex}

\noindent
where we find again the coefficient matrix for $H_0^{(1)}$ in~\eqref{eq:Okmod_H0_1} with the resonance conditions, which coincides with the matrix $H_1^{(3,2)}$. 

Comparing all the systems described by their related Hamiltonian functions at the end of each cascade in both regularisation $r_0$ and $r_1$, we find several examples of Newton polygons with minimal area (i.e.\ $2$), and triangular or diamond-like shape. As already mentioned, the latter is the more minimal, since it has the lowest value for the highest total degree.

The regularisation process iterated twice for the Hamiltonian system associated with $H_{\text{mod}}^{\text{Ok}}$~\eqref{eq:Ok2} can be then summarised with the following diagram: 
{\small
\begin{equation*}
    \begin{tikzpicture}
    \def\lenY{2.5}
    \def\lenX{3.5}
    \def\lenXX{2}
        \path (0,0) node (H) {$\includegraphics[width=0.13\textwidth]{r0_mod_H0.pdf}$} -- ++(\lenX,\lenY) node (H01) {$\includegraphics[width=0.13\textwidth]{r0_H1.pdf}$} -- ++(\lenX,0) node (H011) {$\includegraphics[width=0.13\textwidth]{Okamoto_1.pdf}$} -- ++(\lenXX,0) node (H012) {$\includegraphics[width=0.13\textwidth]{Okamoto_1.pdf}$} -- ++(\lenXX,0) node (H013) {$\includegraphics[width=0.13\textwidth]{r0_H1.pdf}$};
        \path (H) -- ++(\lenX,0) node (H02) {$\includegraphics[width=0.13\textwidth]{r0_mod_H0.pdf}$} -- ++(\lenX,0) node (H021) {$\includegraphics[width=0.13\textwidth]{r0_H1.pdf}$} -- ++(\lenXX,0) node (H022) {$\includegraphics[width=0.13\textwidth]{r0_mod_H0.pdf}$} -- ++(\lenXX,0) node (H023) {$\includegraphics[width=0.13\textwidth]{r0_H1.pdf}$};
        \path (H) -- ++(\lenX,-\lenY) node (H03) {$\includegraphics[width=0.13\textwidth]{r0_H1.pdf}$} -- ++(\lenX,0) node (H031) {$\includegraphics[width=0.13\textwidth]{Okamoto_1.pdf}$} -- ++(\lenXX,0) node (H032) {$\includegraphics[width=0.13\textwidth]{Okamoto_1.pdf}$} -- ++(\lenXX,0) node (H033) {$\includegraphics[width=0.13\textwidth]{r0_H1.pdf}$};
        \node[below=-.3 of H] (a) {\hyperref[eq:H_Ok_mod]{$H^{\text{Ok}}_{\text{mod}}$}}; 
        \node[below=-.3 of H01] (a) {\hyperref[eq:Okmod_H0_1]{$H^{(1)}_{0}$}}; 
        \node[below=-.3 of H02] (a) {\hyperref[eq:Okmod_H0_2]{$H^{(2)}_{0}$}}; 
        \node[below=-.3 of H03] (a) {\hyperref[eq:Okmod_H0_3]{$H^{(3)}_{0}$}};
        \node[below=-.3 of H011] (a) {\hyperref[eq:H1minus]{$H^{(1,-)}_{1}$}}; 
        \node[below=-.3 of H012] (a) {\hyperref[eq:H1plus]{$H^{(1,+)}_{1}$}}; 
        \node[below=-.3 of H013] (a) {\hyperref[eq:H12]{$H^{(1,2)}_{1}\equiv H_0^{(1)}$}};
        \node[below=-.3 of H021] (a) {\hyperref[eq:H21]{$H^{(2,1)}_{1}\equiv H_0^{(1)}$}}; 
        \node[below=-.3 of H022] (a) {\hyperref[eq:H22]{$H^{(2,2)}_{1}$}}; 
        \node[below=-.3 of H023] (a) {\hyperref[eq:H23]{$H^{(2,3)}_{1}\equiv H_0^{(3)}$}};
        \node[below=-.3 of H031] (a) {\hyperref[eq:H3minus]{$H^{(3,-)}_{1}$}}; 
        \node[below=-.3 of H032] (a) {\hyperref[eq:H3plus]{$H^{(3,+)}_{1}$}}; 
        \node[below=-.3 of H033] (a) {\hyperref[eq:H32]{$H^{(3,2)}_{1}\equiv H_0^{(1)}$}};
        \draw[->,thick] (H) -- (H01) node[midway] (r1) {};
        \node[above left] (R1) at (r1) {$r_0$};
        \draw[->,thick] (H) -- (H02) node[midway] (r1) {};
        \node[above ] (R1) at (r1) {$r_0$};
        \draw[->,thick] (H) -- (H03) node[midway] (r1) {};
        \node[above right] (R1) at (r1) {$r_0$};
        \draw[->,thick] (H01) -- (H011) node[midway] (r1) {};
        \node[above] (R1) at (r1) {$r_1$};
        \draw[->,thick] (H02) -- (H021) node[midway] (r1) {};
        \node[above] (R1) at (r1) {$r_1$};
        \draw[->,thick] (H03) -- (H031) node[midway] (r1) {};
        \node[above] (R1) at (r1) {$r_1$};
    \end{tikzpicture}
\end{equation*}
}

\section{\texorpdfstring{The mixed case: quasi-$\text{P}_{\text{IV}}$}{quasiP4}}\label{sec:quasiP4}
We consider the equation introduced in \cite{MDTK1} of the type $\text{P}_{\text{IV}}$
\begin{equation}\label{mixed}
\text{qsi-P}_{\text{IV}}\colon \quad  x''=\frac{(x')^2}{2x}+\frac{5x^5 }{2}+\alpha_4(z) x^4+\alpha_3(z) x^3 +\alpha_2(z) x^2 +\alpha_1(z) x-\frac{\alpha_0^2}{2x},
\end{equation}
with $\alpha_0$ constant, choice that is assumed to consider a polynomial Hamiltonian rather than a rational one. As reminded in the introduction, what generalises the Painlev\'e equations to the quasi-Painlev\'e class is the 
possibility to have not only poles in the solutions as movable singularities, but also branch points. The term ``mixed'' refers to this instance, and the particular case of~\eqref{mixed} presents three different types of singularities, two of which are of square root type, and one is instead a pole, as deepened in \cite{MDTK1}. Following the prescription described in~\eqref{eq:labelling}, this case is addressed as $(2,2,1)$. In the following, we will consider two different Hamiltonian related to~\eqref{mixed}, and in both cases the shortest of the three cascades will be the one associated with the behaviour as a simple pole around the movable singularity. 

The first Hamiltonian  associated with this equation, as studied in \cite{MDTK1}, is 
\begin{equation}\label{H1 mixed}
H_1^{\text{qsi-P}_{\text{IV}}}\big(x(z),y(z);z\big)=\frac{y^2x}{2}-\frac{x^5}{2} -\frac{\alpha_4(z)}{4}x^4-\frac{\alpha_3(z)}{3}x^3 -\frac{\alpha_2(z)}{2}x^2-
\alpha_1(z)x-\alpha_0 y\,, 
\end{equation}
for which the coefficient matrix and the Newton polygon are 
\begin{equation}\label{H1 mixed NP}
H_1^{\text{qsi-P}_{\text{IV}}}=
\left(
\begin{array}{ccc}
 0 & -\alpha_{0} & 0 \\
 -\alpha_{1}(z) & 0 & \frac{1}{2} \\
 -\frac{1}{2}{\alpha_{2}(z)} & 0 & 0 \\
 -\frac{1}{3}{\alpha_{3}(z)} & 0 & 0 \\
 -\frac{1}{4}{\alpha_{4}(z)} & 0 & 0 \\
 -\frac{1}{2} & 0 & 0 \\
\end{array}
\right) \hspace{10ex} \includegraphics[width=.165\textwidth,valign=c]{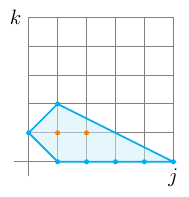} 
\end{equation}
\vspace*{-2ex}

\noindent
with area equal to $5$ and genus $2$.

\noindent
The second Hamiltonian associated with equation \eqref{mixed} is 
\begin{equation}\label{H2 mixed}
H_2^{\text{qsi-P}_{\text{IV}}}\big(x(z),y(z);z\big)=\frac{y^2 x}{2}-y\left(x^3+\frac{\beta_4(z)}{4}x^2+\frac{\beta_3(z)}{3}x+\beta_0(z)\right)-\frac{\beta_2(z)}{2}x^2 -\beta_1(z)x\,,
\end{equation}
where the coefficients $\beta_k(z)$ can be expressed in terms of $\alpha_k(z)$ in~\eqref{H1 mixed} and their derivatives, and vice versa (see~\cite{MDTK1} for the details). 
The related coefficient matrix and Newton polygon  are 
\vspace*{-2ex}
\begin{equation}\label{H2 mixed NP}
H_2^{\text{qsi-P}_{\text{IV}}}=
\left(
\begin{array}{ccc}
 0 & -\beta _0(z) & 0 \\
 -\beta _1(z) & -\frac{1}{3} \beta _3(z) & \frac{1}{2} \\
 -\frac{1}{2} \beta _2(z) & -\frac{1}{4} \beta _4(z) & 0 \\
 0 & -1 & 0 \\
\end{array}
\right) \hspace{10ex} \includegraphics[width=.16\textwidth,valign=c]{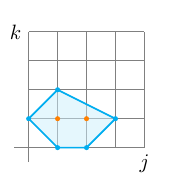} 
\end{equation}
\vspace*{-2ex}

\noindent
with area equal to $7/2$ and therefore smaller compared to~\eqref{H1 mixed}. 
The polynomial regularisation $r_0$ of the Hamiltonian system associated with $H_2^{\text{qsi-P}_{\text{IV}}}$ yields the following constraints for the coefficient functions: 
\begin{equation}
	\beta_0(z) = \beta_0\,,\qquad \beta_2(z)=\beta_2\,,\qquad \beta_4(z) = \beta_{4,0}+\beta_{4,1}z\,.
\end{equation}
The systems obtained from the regularisation procedure are in this case much more complicated than those analysed in the Painlev\'e case.   However, along the shortest cascade (leading to the simple pole) it is possible to generate new Hamiltonian systems for which the Newton polygon is invariant but the coefficient matrix is different. Via the transformation 
\begin{equation}
	x=u,\qquad y=\frac{2\beta_0+u v}{u}
\end{equation}
the regularisation $r_0$ leads to the final Hamitlonian system described by 
\vspace*{-2ex}
\begin{equation}\label{eq:r0_quasiP4}
	H_{2,0}^{\text{qsi-P}_{\text{IV}}}=
	\left(
	\begin{array}{ccc}
		0 & \beta _0 & 0 \\
		-\beta _1(z)-\frac{1}{2} \beta _0 \beta _4(z) & -\frac{1}{3} \beta _3(z) & \frac{1}{2} \\
		-\frac{1}{2} \left(4 \beta _0+\beta _2\right) & -\frac{1}{4} \beta _4(z) & 0 \\
		0 & -1 & 0 \\
	\end{array}
	\right) \hspace{10ex} \includegraphics[width=.16\textwidth,valign=c]{r0_quasiP4_minimal.pdf} 
\end{equation}
\vspace*{-2ex}

\noindent
Two further regularisations, namely $r_1$ and $r_2$ respectively, yield the following coefficients matrices for the Hamiltonian systems at the end of the shortest cascade for each case:
\vspace*{-2ex}
\begin{align}
	H_{2,1}^{\text{qsi-P}_{\text{IV}}}&=
	\left(
	\begin{array}{ccc}
		0 & -\beta _0 & 0 \\
		-\beta_1(z) & -\frac{1}{3}\beta _3(z) & \frac{1}{2} \\
		-\frac{1}{2} \beta _2 & -\frac{1}{4}
		\left(\beta _{4,0}+z \beta _{4,1}\right) & 0 \\
		0 & -1 & 0 \\
	\end{array}
	\right) \hspace{10ex} \includegraphics[width=.16\textwidth,valign=c]{r0_quasiP4_minimal.pdf}  \\[.5ex]
    	H_{2,2}^{\text{qsi-P}_{\text{IV}}}&=
	\left(
	\begin{array}{ccc}
		0 & \beta _0 & 0 \\
		-\frac{1}{2}\beta_0 (\beta_{4,0}+z\beta_{4,1})-\beta _1(z) & -\frac{1}{3}\beta _3(z) & \frac{1}{2} \\
		-2\beta_0-\frac{1}{2}\beta_2 & -\frac{1}{4} \left(\beta _{4,0}+z \beta
		_{4,1}\right) & 0 \\
		0 & -1 & 0 \\
	\end{array}
	\right) \hspace{5ex} \includegraphics[width=.16\textwidth,valign=c]{r0_quasiP4_minimal.pdf} 
\end{align}
\vspace*{-2ex}

\noindent
with the same polygon as in~\eqref{eq:r0_quasiP4}. 
Note that for all these Hamiltonians, one of the variables in which it is determined, satisfies the equation~\eqref{mixed}. The final charts along other cascades are much more complicated and therefore it is much more complicated to derive the possible Hamiltonian formulation. In that case the Hamiltonian function to be found would not be Hamiltonian with respect to the canonical symplectic form $u_{\text{fin}}\wedge v_{\text{fin}}$, but rather to the quasi-canonical symplectic form (see appendix \ref{subapp:intermediate}).  
It is then reasonable to consider the Newton polygon in~\eqref{eq:r0_quasiP4} as the simplest possible in this case. 

The iterative polynomial regularisation of the Hamiltonian system associated with $H_{1}^{\text{qsi-P}_{\text{IV}}}$ as in~\eqref{H1 mixed} along some of the cascades, leads to the same simple version of the Newton polygon as in~\eqref{H2 mixed NP}. After the first regularisation $r_0$, one of the end of cascades is obtained via the transformation of the original variables $(x,y)\mapsto (u,v)$
\begin{equation}
    x=\frac{1}{u}\,,\qquad v=\frac{P(u(z),v(z);z)}{u^2}\,,
\end{equation}
where the function $P$ has the form 
\begin{equation}
    \begin{split}
        P&=1+u^2\! \left(\frac{\alpha_{3}(z)}{3}-\frac{\alpha_{4}(z)^2}{32}\right)+ u^3 \!\left(\alpha_{0}+\frac{\alpha_{2}(z)}{2}-\frac{1}{12} \alpha_{3}(z) \alpha_{4}(z)-\frac{\alpha_{4}'(z)}{8} +\frac{\alpha_{4}(z)^3}{128}\right)\\[1ex]
        &~~+u^4 \!\left(\alpha_{1}(z)+\frac{1}{32} \alpha_{3}(z) \alpha_{4}(z)^2-\frac{\alpha_{3}(z)^2}{18}-\frac{1}{128} \alpha_{4}(z)^4\right) +u^5 \!\left(v-\frac{1}{12} \alpha_{4}(z) \alpha_{3}'(z)\right) \\[1ex]
        &~~+u^8 \alpha_{4}(z)\left(\frac{1}{24} \alpha_{2}(z)  \alpha_{3}'(z)-\frac{1}{32}  \alpha_{3}'(z) \alpha_{4}'(z)\right).
    \end{split}
\end{equation}
The regularising conditions on the coefficients are 
\begin{equation}
    \alpha_4''(z)=0\,, \qquad \big(192\, \alpha_2(z)-32\,\alpha_3(z)\alpha_4(z)+3\,\alpha_4(z)^3\big)'=0 \,,
\end{equation}
so that the coefficient functions can be expressed as 
\begin{equation}
    \alpha_4(z)=\alpha_{4,0}+\alpha_{4,1}z\,, \qquad \alpha_2(z)=\alpha-3(\alpha_{4,0}+\alpha_{4,1}z)^3+32\alpha_3(z)(\alpha_{4,0}+\alpha_{4,1}z)\,,
\end{equation}
with $\alpha_{4,0},\alpha_{4,1}$ and $\alpha$ constants. 

Proceeding with the regularisation $r_1$ of the system in coordinates $(u,v)$, we get the final chart with the transformation of variables 
\begin{equation}
    u=\frac{1}{u_1}\,,\qquad v=u_1\, P_1(u_1,v_1,z)\,,
\end{equation}
where the function $P_1$ is 
\begin{equation}
    \begin{split}
        P_1&= u_{1} \!\left(u_{1} v_{1}-\frac{\alpha}{384}-\alpha_{0}+\frac{\alpha_{4,1}}{8}\right)-\alpha_{1}(z)+\frac{\alpha_{3}'(z)}{3}-\frac{1}{96} \alpha_{3}(z) \left(\alpha_{4,0}+\alpha_{4,1} z\right)^2+\frac{\alpha_{3}(z)^2}{18} \\[1ex]
        &~~+ \frac{1}{6144}(\alpha_{4,0}+\alpha_{4,1}z) \left(4\alpha+3(\alpha_{4,0}+\alpha_{4,1}z)^3-576\,\alpha_{4,1}\right)\,. 
    \end{split}
\end{equation}
The corresponding Hamiltonian for this final system after the regularisation $r_1$ takes the form 
\begin{equation}\label{eq:r1_quasiP4}
	H_{1,1}^{\text{qsi-P}_{\text{IV}}}=
	\left(
\begin{array}{ccc}
 0 & -\alpha _0 & 0 \\[.5ex]
 h(z) &
   \frac{1}{3}\alpha _3(z)-\frac{1}{32} \left(\alpha_{4,0}+\alpha_{4,1}z\right)^2 &
   \frac{1}{2} \\[.5ex]
 \frac{1}{8}\alpha _{4,1}-\frac{1 }{384}\alpha-\alpha _0 & \frac{1}{4} \left(\alpha_{4,0}+\alpha_{4,1}z\right) & 0 \\[.5ex]
 0 & 1 & 0 \\
\end{array}
\right) \hspace{2ex} \includegraphics[width=.16\textwidth,valign=c]{r0_quasiP4_minimal.pdf} 
\end{equation}
\vspace*{-2ex}

\noindent
where the function $h$ is 
\begin{equation}
    \begin{split}
    h(z)&= -\alpha_{1}(z)+\frac{\alpha_{3}(z)^2}{18}-\frac{1}{96} \alpha_{3}(z) (\alpha_{4,0}+\alpha_{4,1} z)^2+\frac{1}{2048}(\alpha_{4,0}+\alpha_{4,1} z)^4 +\frac{ \alpha_{3}'(z)}{3}  \\[1ex]
    &~~~~- \frac{1}{16}(\alpha_{4,0}+\alpha_{4,1} z)(4 \alpha_{40}+\alpha_{4,1})\,.
    \end{split}
\end{equation}
Tracking back the transformations of variables, the variable $u_1$ in~\eqref{eq:r1_quasiP4} corresponds to the original variable~$x$, therefore it satisfies the same differential equation as in~\eqref{mixed}, subject to the resonance conditions on the coefficient functions. It is worth noticing that the Newton polygon as in~\eqref{eq:r1_quasiP4} appears in other final charts in the iterative regularisation process. 
Moreover, we observe that the genus of the Newton polygon might increase.

Therefore, the iterative regularisation at the second step applied to $H_1^{\text{qsi-P}_{\text{IV}}}$ shows a reduction of the area of the original Newton polygon (from $5$ to $7/2$). The Hamiltonian $H_2^{\text{qsi-P}_{\text{IV}}}$ then is to be considered to have a more ``minimal'' form compared to $H_1^{\text{qsi-P}_{\text{IV}}}$. At the moment, we do not know whether it is possible to find a different Hamiltonian function for the system related to~\eqref{mixed} such that the corresponding Newton polygon has an area smaller than 7/2.

\section{Open questions}\label{sec:open}
As already mentioned, the theory for systems with the quasi-Painlev\'e property has not been established yet. One important problem is associated with the classification of this type of systems, that goes through the identification of minimal cases. 
The iterative polynomial regularisation might provide a useful tool for this problem. In particular, it would be interesting to review the results in~\cite{MDTK2} with this machinery, to see what kind of systems appear in the orbits of different regularisations. In the case of the quasi-Painlev\'e systems, there is no uniqueness result about their global Hamiltonian structure. Therefore, finding ``minimal'' Hamiltonians corresponding to the instances of the Newton polygons with minimal areas could be effectively used for the classification of quasi-Painlev\'e equations and then to solve the quasi-Painlev\'e equivalence problem (i.e.\ to identify whether a given equation or system can be reduced to some standard form). The value of highest total degree per polynomial should be also taken into account to establish the minimal version of the Hamiltonian. 

A satisfying theory for systems of quasi-Painlev\'e type should provide a generalisation of different features of systems of Painlev\'e type. 
Systems with quasi-Painlev\'e property might be interesting in the context of reductions of integrable and non-integrable PDEs, following the classical problem of reductions (see e.g.~\cite{Hone2009}). For instance, $\text{P}_{\text{II}}$ is known to arise from a scaling reduction of the Korteweg-de Vries (KdV) equation 
\begin{equation}\label{eq:kdv}
    u_t-3(n+1)u^n\,u_x+u_{xxx}=0\,,
\end{equation}
with $n=1$, as well as in the case of the modified KdV ($n=2$), where the nonlinear term is squared. It is possible to generalise the scaling reduction to the problem of the generalised KdV, for $n>2$. In particular, for $n=4$ the re-scaled equation takes the form 
\begin{equation}\label{eq:kdv_1}
    t^{1/6}\, u_t+\sqrt{x}\,(15\, u^4\, u_x+u_{xxx})=0\,,
\end{equation}
from which by introducing the new variable $y$ as 
\begin{equation}
    u(t,x) = u(t,x)=-\frac{1}{t^{1/6}} \,\, y\!\left(\frac{x}{t^{1/3}}\right) \,,
\end{equation}
and integrating once, we obtain a specific case of the quasi-Painlev\'e equation of type II: 
\begin{equation}
    \text{quasi-P}_{\text{II}}\colon\quad y''(z)=-3\,y(z)^5+\frac{\sqrt{z}}{3}\,y(z)-\alpha \,, 
\end{equation}
with $z=x/t^{1/3}$. The quasi-Painlev\'e equations thus might inherit some structures from the associated PDEs and this might give new applications of the quasi-Painlev\'e equations. 

Lastly, we note something about the solutions to the mixed case for quasi-Painlev\'e equations in analogy with Painlev\'e equations.  
The mixed equation~\eqref{mixed} admits solutions which solve the first order differential equation 
\begin{equation}
    x'=\sum_{k=0}^3 a_k(z)\, x^k\,, 
\end{equation}
where the functions $a_k(z)$ for $k=0\,, \dots, 3$ can be expressed in terms of the coefficients $\alpha_k(z)$ appearing in~\eqref{mixed}. Moreover, there are certain conditions on the coefficient functions $\alpha(z)$ and their derivatives, which are quite lengthy so we omit them. This resembles the so-called Riccati solutions for the Painlev\'e equations. If the equation~\eqref{mixed} is expressed in terms of elements in $x$ with more negative powers (e.g.~up to $x^{-3}$), then one could have found solutions satisfying 
\begin{equation}
    x'=\sum_{k=-1}^3 a_k(z)\, x^k\,.
\end{equation} 
This deserves further study to build the theory of quasi-Painlev\'e equations similar to the one for the Painlev\'e equations.

\section*{Acknowledgements}
The work of GF is part of the project ``ERDF A way of making Europe'' and the project PID2021-124472NB-I00  funded by MICIU/AEI/10.13039/501100011033.

\begin{appendices}
\section{Regularisation procedure}\label{app:regularisation}
Let us recall the basic steps to regularise a system of the type 
\begin{equation}\label{eq:original}
    \begin{cases}
        x'= P(x,y;z) \\
        y'= Q(x,y;z)
    \end{cases},
\end{equation}
with $x=x(z)$, $y=y(z)$ and $P$ and $Q$ polynomial functions, following the approach in~\cite{Okamoto1979,Sakai2001}.   
\subsection{\texorpdfstring{Compactification of $\mathbb{C}^2$}{comp}}
Firstly, the phase space of the system $\big(x(z),y(z)\big) \in \mathbb{C}^2$, is compactified  to include points at infinity. The compactification of $\mathbb{C}^2$ can be performed in different ways, typically as $\mathbb{P}^2$ or $\mathbb{P}^1\times \mathbb{P}^1$ or, more in general, a Hirzebruch surface. In the paper, we will be using the compactification $\mathbb{P}^1\times \mathbb{P}^1$, given by the glueing of four affine charts: 
\begin{equation}\label{eq:comp}
\includegraphics[width=.25\textwidth,valign=c]{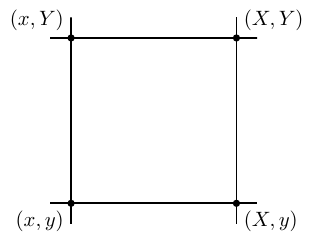} \qquad \begin{aligned} 
    &\mathbb{P}^1 \times \mathbb{P}^1 \simeq  \mathbb{A}^2_{(x,y)} \cup \mathbb{A}^2_{(X,y)} \cup \mathbb{A}^2_{(x,Y)} \cup \mathbb{A}^2_{(X,Y)}\,, 
    \\[1ex] 
    &X = \frac{1}{x}\,,\quad  Y = \frac{1}{y}\,. 
    \end{aligned}
\end{equation}
In each affine chart, the system is represented as a pair of rational differential equations in the corresponding variables: 
\begin{equation}\label{eq:P1P1}
    \begin{cases}
        x'= P_1(x,Y;z) \\
        Y'= Q_1(x,Y;z)
    \end{cases}, \qquad \begin{cases}
        X'= P_2(X,Y;z) \\
        Y'= Q_2(X,Y;z)
    \end{cases}, \qquad \begin{cases}
        X'= P_3(X,y;z) \\
        y'= Q_3(X,y;z)
    \end{cases},
\end{equation}
where now the functions $P_k, Q_k$ for $k=1,2,3$ are in general rational functions. 

\subsection{Blow-up transformations}
We analyse the system~\eqref{eq:original} together with the systems in  the other affine charts in~\eqref{eq:P1P1}. We identify the so-called 
base points by looking for (finitely many) points of indeterminacy labelled as
$p_j$ ($j=1,\dots,N$), for which the right hand side of the rational differential equations becomes~$0/0$. These are points of coalescence of infinitely many integral curves. In order to resolve the indeterminacy, and with the aim of distinguishing each curve passing through such a point, we introduce a blow-up transformation of the surface $\mathbb{P}^1 \times \mathbb{P}^1$, such that the point is substituted by a complex projective line $\mathbb{P}^1$. In this way, each of the previously coalescing curves intersects the line $\mathbb{P}^1$ in one point only, as depicted heuristically here:
\begin{equation*}
    \includegraphics[width=0.5\linewidth]{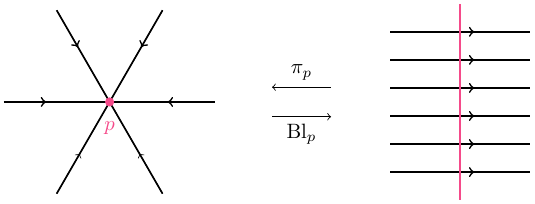}
\end{equation*}
To completely describe the line $\mathbb{P}^1$ emerging in place of the indeterminacy point $p$ (visible  e.g.\ in the chart $(x,y)$ with coordinates $p\colon (x=a,y=b)$), one needs to introduce two new charts with coordinates $(u,v)$ and $(U,V)$ via birational transformations
\begin{equation}\label{eq:blowup}
    \begin{cases}
        u = x-a \\[1ex]
        v = \dfrac{y-b}{x-a} 
    \end{cases}\,,  \hspace{5ex} \begin{cases}
        U = \dfrac{x-a}{y-b} \\[1ex]
        V = y-b
    \end{cases} \,. 
\end{equation}
The line isomorphic to $\mathbb{P}^1$ emerging from this transformation is referred to as the exceptional curve~$E$ satisfying
\begin{equation}\label{eq:exc_curve}
    E = \{u = 0\} \cup \{ V = 0 \}\,.  
\end{equation}
The system of differential equations in which the indeterminacy point is visible is then transformed in the new two charts, giving rise to two rational differential systems in the coordinates $(u,v)$ and $(U,V)$ respectively. These systems may have new points of indeterminacy, at which we apply again the blow-up transformation. The finite sequence of these transformations at points $q_i^{(j)}$ gives rise to the so-called cascade. We shall also use the upper index $(j)$ to indicate the $j$-th cascade emerging from the point $p_j$ in the original system. 

In the $j$-th cascade the convention for the change of variables (omitting the upper index) $(u_i,v_i) \mapsto (u_{i+1},v_{i+1}) \cup (U_{i+1},V_{i+1})$ to perform the blow-up in the chart $(u_i,v_i)$ at the point $q_i\colon(u_{i},v_i)=(a_i,b_i)$ is  
\begin{equation}
    \begin{cases}
        u_{i+1} = u_{i}-a_i \\[1ex]
        v_{i+1} = \dfrac{v_i - b_i}{u_i - a_i}
    \end{cases}, \qquad \begin{cases}
        U_{i+1} = \dfrac{u_{i} - a_i}{v_i - b_i} \\[1.5ex]
        V_{i+1} = v_i + b_i 
    \end{cases}.
\end{equation}
Every time a blow-up is performed, an additional exceptional curve emerges, represented as the set $E_{i+1}=\{u_{i+1}=0\} \cup \{V_{i+1}=0\}$, adapting the formulation in~\eqref{eq:exc_curve}. This procedure is repeated until the system represented in both final charts $(u_{\text{fin}},v_{\text{fin}})$ and $(U_{\text{fin}},V_{\text{fin}})$ is not affected by further indeterminacies. At the end of the procedure the system is said regularised if it is regular on the last exceptional curve in one of the final charts. Hence, the regularisation process gives rise to a chain of finitely many consecutive birational transformations of variables, e.g.\ 
\begin{equation}\label{eq:chain}
    (x,y) ~\mapsto~ (u_1,v_1) ~\mapsto~ \dots~ \mapsto ~(U_k,V_k)~ \mapsto ~ \dots~ \mapsto ~(u_{\text{fin}},v_{\text{fin}}) \,. 
\end{equation}
We shall also report, if any, the resonance conditions arising in the regularisation process, i.e.\ (differential) constraints on the coefficient functions appearing in the systems which make the system in one of the final charts regular\footnote{
Finding the regularisation conditions 
is equivalent to fixing the resonance conditions on the coefficients in the series expansion of solutions in the (quasi)-Painlev\'e test (see~\cite{Hone2009,GF_AS1,GF_AS2,TKGF,MDTK1}).}. 
The final system (e.g.\ in the chart $(u_{\text{fin}},v_{\text{fin}})$) can be solved in terms of the coordinates on the last exceptional curve ($E_{\text{fin}}=\{u_{\text{fin}=0}\}$)  implicitly in the variables $(z(u_{\text{fin}}),v_{\text{fin}}(u_{\text{fin}}))$, with initial condition~$(z_*,h)\in \mathbb{A}^2$ and arbitrary $h$. The series expressions are then inverted, so that the variables $(u_{\text{fin}},v_{\text{fin}})$ are given as functions of $z-z_*$. Tracking back all the regularising transformations in~\eqref{eq:chain}, one can derive the behaviour of the original variables~$(x,y)$, in the neighbourhood of the movable singularity $z_*$, as a function of $z-z_*$, giving rise to either Laurent series~\eqref{eq:laurent} or Puiseux series~\eqref{eq:puiseux}.

\subsection{Intermediate transformations for polynomial regularisation of Hamiltonian systems}\label{subapp:intermediate}

Suppose that our original system~\eqref{eq:original} is Hamiltonian with respect to the (standard) symplectic 2-form $dx \wedge dy$. For the case of Hamiltonian systems associated with Painlev\'e equations, the global Hamiltonian structure was studied in~\cite{Takano97,Takano99}. The authors constructed global symplectic atlas for the systems, given by glueing the affine charts obtained after the final blow-up in every cascade with the original one. In the final chart of each cascade, the system is described by a polynomial Hamiltonian with respect to the $2$-form 
\begin{equation}\label{eq:final_var}
    dx \wedge dy = \ell \, du_{\text{fin}} \wedge dv_{\text{fin}}  \qquad \text{or} \qquad dx \wedge dy = \ell\, dU_{\text{fin}}  \wedge dV_{\text{fin}}   \,,
\end{equation}
with $|\ell|$ the ramification number, $|\ell|=1$ for $\text{P}_{\text{II}}$ -- $\text{P}_{\text{VI}}$ and $|\ell|=2$ for $\text{P}_{\text{I}}$~\cite{IwasakiOkada}. This construction was generalised in the quasi-Painlev\'e case in~\cite{GF_AS1,GF_AS2}, where the $2$-form in the final chart of each cascade is not any more canonical, and given by
\begin{equation}\label{eq:quasi_final_var}
    dx \wedge dy = \ell\,u^{k-1} du_{\text{fin}} \wedge dv_{\text{fin}}  \qquad \text{or} \qquad dx \wedge dy = \ell\,V^{k-1}  \,dU_{\text{fin}}  \wedge dV_{\text{fin}}   \,,
\end{equation}
where $\ell$ is the same as above, and $1/k$ is the branching order in the Puiseux series~\eqref{eq:puiseux}. 

In order to obtain these expressions for the $2$-forms in the final charts, one can include (one of the) two types of intermediate changes of variables in the cascades. The first intermediate change $(u,v) \mapsto (\,\widetilde{u}\,,\widetilde{v}\,)$ (or $(U,V) \mapsto (\,\widetilde{U},\widetilde{V}\,)$) is
\begin{equation}\label{eq:generic_twist}
    u = \widetilde{u} \,, \quad 
    v = \dfrac{1}{\widetilde{v}} 
\qquad \text{ or } \qquad 
    U = \dfrac{1}{\widetilde{U}}  \,, \quad 
    V = \widetilde{V} \,,
\end{equation}
and the second $(u,v) \mapsto (\,\widehat{u}\,,\widehat{v}\,)$ (or $(U,V) \mapsto (\,\widehat{U},\widehat{V}\,)$) is the $\ell$-fold covering
\vspace*{-2ex}
\begin{equation}\label{eq:generic_covering}
    u = \dfrac{\widehat{u}}{\widehat{v}} \,, \quad 
    v = \widehat{v}^{\,\ell}  
  \qquad \text{ or } \qquad 
    U = \widehat{U}^{\,\ell} \,, \quad 
    V = \dfrac{\widehat{V}}{\widehat{U}} \,. 
\end{equation}
In the paper, we will mark the variables with either $\sim$ or $\,\,\widehat{\,}\,\,$ in the case we use either~\eqref{eq:generic_twist} or~\eqref{eq:generic_covering} respectively.

\end{appendices}

\bibliography{thebiblio}
\bibliographystyle{style_alpha}

\end{document}